\documentclass[twocolumn,prc,aps,superscriptaddress]{revtex4}

\usepackage{epsfig}
\usepackage{graphicx}% Include figure files
\usepackage{dcolumn}% Align table columns on decimal point
\usepackage{bm} % bold math
\usepackage{epstopdf}
\usepackage{amsmath}
\usepackage{color}
\usepackage{hyperref}

\newcommand{\beq}{\begin{equation}}
\newcommand{\eeq}{\end{equation}}
\newcommand{\beqa}{\begin{eqnarray}}
\newcommand{\eeqa}{\end{eqnarray}}
\newcommand{\la}{\langle}
\newcommand{\ra}{\rangle}

\newcommand{\tcb}{\textcolor{blue}}

\begin{document}

\title{Nuclear excitation cross section of $^{229}$Th via inelastic electron scattering}

\author{Hanxu Zhang}
\affiliation{Graduate School, China Academy of Engineering Physics, Beijing 100193, China}

\author{Wu Wang}
\affiliation{Beijing Computational Science Research Center, Beijing 100193, China}

\author{Xu Wang}
\email{xwang@gscaep.ac.cn}
\affiliation{Graduate School, China Academy of Engineering Physics, Beijing 100193, China}

\date{\today}

\begin{abstract}
Nuclear excitation cross section of $^{229}$Th from the ground state to the low-lying isomeric state via inelastic electron scattering is calculated, on the level of Dirac distorted wave Born approximation. With electron energies below 100 eV, inelastic scattering is very efficient in the isomeric excitation, yielding excitation cross sections on the order of 10$^{-27}$ to 10$^{-26}$ cm$^2$. Systematic analyses are presented on elements affecting the excitation cross section, including the ion-core potential, the relativistic effect, the knowledge of the reduced nuclear transition probabilities, etc. 
\end{abstract}

\maketitle

\section{Introduction}

%\node[options] (id) at position {label};

The nucleus of $^{229}$Th has a unique low-lying isomeric state with an energy of only about 8.28 eV above the nuclear ground state. Historically, the existence of this isomeric state was first suggested by Kroger and Reich in 1976 \cite{kroger1976}. The energy was estimated to be below 100 eV \cite{kroger1976}, $1\pm 4$ eV \cite{reich1990energy}, $3.5\pm 1.0$ eV \cite{helmer1994excited}, $7.6\pm 0.5$ eV \cite{beck2007energy}, and recently $8.28\pm 0.17$ eV \cite{seiferle2019energy}, with progresses of experimental techniques. Being the only known nuclear state on the 1-eV order of magnitude (the next lowest nuclear state being the isomeric state of $^{235}$U, which has an energy of 76 eV \cite{browne2014}), the $^{229}$Th nucleus has attracted much attention in recent years for its potential applications in nuclear optical clocks \cite{pauli1924, peik2003nuclear, rellergert2010constraining, campbell2012single, peik2015nuclear}, nuclear lasers \cite{Tkalya-11}, checking temporal variations of fundamental constants \cite{Flambaum-06, Berengut-09, Fadeev-20}, etc.

The isomeric state can be obtained from $\alpha$ decay of $^{233}$U, with $2\%$ of the resultant $^{229}$Th nuclei in the isomeric state. The disadvantages of this ``natural" way of obtaining the isomeric state include: (i) Low efficiency due to the long half-time ($1.6\times 10^5$ years) and the low branching ratio of the decay reaction. One can estimate that every $3.6\times 10^{14}$ $^{233}$U nuclei generate a single $^{229}$Th nucleus in the isomeric state in one second. (ii) The $^{229}$Th nuclei are left with a recoil energy of 84 keV into random directions and a variety of ionic states. Alternatively, the isomeric state can be obtained from $\beta$ decay of $^{229}$Ac \cite{Verlinde-19}, suffering however from the low production yield of $^{229}$Ac. 

Controllable and efficient approaches of generating the isomeric state are therefore highly desirable. Direct optical excitation from the nuclear ground state using vacuum ultraviolet light is logically most straightforward. The idea is to radiate the $^{229}$Th nuclei for some time, and then to detect the subsequent fluorescence to see whether it has the desired half-life decay feature. Several experimental attempts have been made without success \cite{jeet2015results, yamaguchi2015experimental, stellmer2018attempt, von2017laser}. Possible reasons include inaccurate knowledge of the isomeric energy, competing fluorescence signals from the electrons with similar photon energies, competition with nonradiative channels, etc. An indirect optical excitation scheme was demonstrated experimentally by Masuda et al. \cite{masuda2019x}, who use narrow-band 29 keV synchrotron radiations to excite the $^{229}$Th nuclei from the ground state to the second excited state which then decays preferably into the isomeric state \cite{Tkalya-00}. The probability for a single $^{229}$Th nucleus to be excited to the isomeric state is estimated experimentally to be on the order of $10^{-11}$ per second.
Excitation approaches exploiting the coupling between the nuclear and the electronic degrees of freedom have also been proposed. Schemes based on electronic bridge processes have been considered for various ionic states or doped-crystal systems \cite{Tkalya-92, porsev2010electronic, porsev2010excitation, bilous2018laser, dzyublik2020excitation, bilous2020electronic, Nickerson-20}. Proposals have also been made using inelastic scattering of the electron \cite{tkalya2020excitation} or laser-driven electron recollision \cite{wang2021exciting, wang2021strong, Wang2022}.

In the current article we consider isomeric excitation of $^{229}$Th via inelastic electron scattering. Although electronic excitation of nuclei is a common tool in nuclear physics \cite{Hofstadter-1953, barber1962inelastic, rosen1967generalized, theissen1972spectroscopy, donnelly1975electron, Bertrand-1976, Taylor-1991, harston1999mechanisms, Uberall-2012}, the electron energies are usually high ($> 1$ MeV) due to typical nuclear energy scales, and few attentions have been paid to low electron energies. In 2020 Tkalya showed that inelastic scattering of low-energy electrons can be very efficient in exciting the $^{229}$Th nucleus \cite{tkalya2020excitation}. For electron energies below 100 eV, the isomeric excitation cross sections are shown to be on the order of 10$^{-27}$ to 10$^{-26}$ cm$^2$. 
The goal of the current article is twofold. One is to provide an independent derivation and calculation on the same subject. We find that Tkalya's results have a couple of minor errors, including larger by an overall factor of two and a confusion of the nuclear transition direction in the reduced transition probabilities. 
The other goal is to provide a detailed analysis on elements affecting the excitation cross section. These elements include the ion-core potential, the relativistic effect, the reduced transition probabilities, etc. We believe that such an analysis is helpful in understanding the electronic excitation process and the robustness of the obtained cross sections.   

This article is structured as follows: 
In Sec. \ref{II} we present a theoretical framework for nuclear excitation via inelastic electron scattering. The theoretical framework is on the level of Dirac distorted wave Born approximation.  
In Sec. \ref{III} we present numerical results and discuss dependency of the excitation cross section on several factors, including the existence or absence of the ion-core potential, the relativistic effect, the ionic state, and the reduced transition probabilities. 
A conclusion is given in Sec. \ref{IV}.

\section{Method}\label{II}

Consider an electron, coming from infinity with asymptotic energy $E_i$, scatters off a $^{229}$Th atom or ion. We consider the channel that the nucleus gets excited from the ground state to the isomeric state and the electron leaves with an asymptotic final energy $E_f = E_i - 8.28$ eV. We neglect higher-order channels in which both the nucleus and the atomic (ionic) electron cloud get excited. The atomic (ionic) electron cloud just supplies a static mean-field potential for the scattering electron.

The differential excitation cross section can be obtained by Fermi's golden rule (atomic units are used unless otherwise specified, $\hbar=m_e=e=1$)
\beq\label{eq_FermiGR}
\frac{d\sigma}{d\Omega}=\frac{2\pi}{v_i}\rho(E_f)|\la f | H_\text{int} | i \ra|^2 ~,
\eeq
where $\Omega$ is the solid angle of the outgoing direction, $v_i=p_ic^2/E_i$ is the asymptotic incoming speed, 
$\rho(E_f)=p_fE_f/\left(8\pi^3c^2\right)$ is the density of the final states, 
$E_{i,f}=\sqrt{p_{i,f}^2c^2+m_e^2c^4}$ is the energy of the initial or the final state. Including electron spin, averaging over initial states and summing over final states, Eq. (\ref{eq_FermiGR}) can be written as
\beq\label{eq_dif_cross1}
\frac{d\sigma}{d\Omega}=\frac{E_iE_f}{4\pi^2c^4}\frac{p_f}{p_i}\frac{1}{2(2I_i+1)}
\sum_{M_iM_f}\sum_{\nu_i\nu_f}|\la f | H_\text{int} | i \ra|^2 ~.
\eeq
Here $I_{i,f}$ and $M_{i,f}$ are the total angular momentum and the magnetic quantum number of the initial or the final state of the nucleus. $\nu=\pm 1/2$ represents spin up or spin down of the electron. Introduce spinor $\chi^\nu$:
\beq
\chi^{1/2}=\binom{1}{0}\ \ \ \rm{and}\ \ \ \chi^{-1/2}=\binom{0}{1} ~, \label{e.chinu}
\eeq 
which are to be used later.

In the matrix element $\la f | H_\text{int} | i \ra$, the initial state and the final state are
\beqa
| i \ra &=&  | I_i M_i \ra \otimes | \phi_i \ra \otimes | 0 \ra~, \\
| f \ra &=&  | I_f M_f \ra \otimes | \phi_f \ra \otimes | 0 \ra~.
\eeqa
That is, the state of the total scattering system is the product of the states of the nucleus ($| I M \ra$), 
of the scattering electron ($| \phi \ra$), and of the radiation field ($| n \ra$). 
The Hamiltonian of the system can be written as
\beq
H = H_n + H_e + H_\text{rad} + H_\text{int}~,
\eeq
consisting of the Hamiltonians of the nucleus $H_n$, of the scattering electron $H_e$, of the radiation field $H_\text{rad}$, and of the interaction $H_\text{int}$. The latter is given by
\beq
\begin{split}
H_\text{int} =  
 &- \frac{1}{c} \int [\bm{j}_n(\bm{r}) + \bm{j}_e(\bm{r})] \cdot \mathbf{A}(\bm{r}) d\tau \\
 &+ \int \frac{\rho_n(\bm{r}) \rho_e(\bm{r'})}{|\bm{r}-\bm{r'}|} d\tau d\tau'  
\end{split}
\eeq
where the first integral is the couplings between the nuclear current density $\bm{j}_n$ and the electron current density $\bm{j}_e$ with the vector potential $\mathbf{A}$ of the radiation field.
The second integral is the Coulomb interaction between the nucleus and the electron, with $\rho_n$ and $\rho_e$ being the charge density operator of the nucleus and of the scattering electron, respectively.
The vector potential of the radiation field can be expanded in multipole components as
\beqa
\mathbf{A}(\bm{r}) = \sum_{\lambda \mu q} &[&  a(E\lambda,\mu,q)\bm{A}(E\lambda,\mu,q) \nonumber\\
&& + a(M\lambda,\mu,q)\bm{A}(M\lambda,\mu,q)+h.c. ] ~.
\eeqa
In the above expression, $\lambda,\mu,q$ are the angular momentum quantum number, magnetic quantum number, and wave number, respectively,  
and
\beqa
\bm{A}(E\lambda,\mu,q) &=& \sqrt{\frac{8\pi c^2}{\lambda (\lambda+1)R}}\nonumber
\nabla \times \bm{L} \left[j_{\lambda}(qr)Y_{\lambda \mu}(\theta, \phi) \right] \label{eq_AE} \\
\\
\bm{A}(M\lambda,\mu,q) &=& i\sqrt{\frac{8\pi c^2 q^2}{\lambda (\lambda+1)R}}
\bm{L} \left[j_{\lambda}(qr)Y_{\lambda \mu}(\theta, \phi) \right]\label{eq_AM}
\eeqa
where $R$ is the radius of the spherical volume under consideration, 
$\bm{L}$ is the angular momentum operator, 
$j_{\lambda}(qr)$ is a spherical Bessel function, 
and $Y_{\lambda \mu}$ is the spherical harmonics.
The expansion coefficient $a$ and its conjugate are the operators for photon annihilation and creation. The matrix elements of these operators are
\beq
\la n | a | n+1 \ra=\la n+1 | a^\dagger | n \ra= \sqrt{ \frac{n+1}{2qc} }
\eeq
where $|n\ra$ represents a number state with $n$ photons. 

The matrix element of $H_\text{int}$ can be derived into the following form \cite{alder1958errata}
\beq\label{eq_Hfi}
\begin{split}
&\la f|H_\text{int}| i\ra =\sum_{\lambda\mu}\frac{4\pi}{2\lambda+1}(-1)^{\mu}\\
&\times \left\{\la \phi_f |\mathcal{N} (E\lambda,\mu)|\phi_i\ra\la I_fM_f|\mathcal{M} (E\lambda,-\mu)| I_iM_i\ra \right.\\
&\left. - \la \phi_f|\mathcal{N} (M\lambda,\mu)|\phi_i\ra\la I_fM_f|\mathcal{M} (M\lambda,-\mu)| I_iM_i\ra \right\},
\end{split}
\eeq
where $\mathcal{M} (\mathcal{T} \lambda,\mu)$ and $\mathcal{N} (\mathcal{T} \lambda,\mu)$ are the electric $(\mathcal{T}=E)$ or magnetic $(\mathcal{T}=M)$ multipole transition operators of the nucleus and of the electron, respectively:
\beq\label{eq_nu_matrix_E} 
\begin{split}
\mathcal{M} (E \lambda, \mu) &= \frac{(2\lambda+1)!!}{\kappa^{\lambda+1}c(\lambda+1)}\\
&\times \int \bm{j}_n \cdot \nabla \times \bm{L} \left[j_{\lambda}(\kappa r)Y_{\lambda \mu}(\theta, \phi) \right]d\tau
\end{split}
\eeq
\beq\label{eq_nu_matrix_M}
\begin{split}
\mathcal{M} (M \lambda,\mu) &=\frac{-i(2\lambda+1)!!}{\kappa^{\lambda}c(\lambda+1)}\\
&\times \int \bm{j}_n \cdot \bm{L} \left[j_{\lambda}(\kappa r)Y_{\lambda \mu}(\theta, \phi) \right] d\tau
\end{split}
\eeq
\beq\label{eq_el_matrix_E}
\begin{split}
\mathcal{N} (E\lambda,\mu) &= \frac{i\kappa^{\lambda}}{c\lambda (2\lambda-1)!!}\\
&\times \int \bm{j}_e \cdot \nabla \times \bm{L} \left[h_{\lambda}^{(1)}(\kappa r)Y_{\lambda \mu}(\theta, \phi) \right] d\tau
\end{split}
\eeq
\beq\label{eq_el_matrix_M}
\begin{split}
\mathcal{N} (M \lambda,\mu) &= \frac{\kappa^{\lambda+1}}{c\lambda (2\lambda-1)!!}\\
&\times \int \bm{j}_e \cdot \bm{L} \left[h_{\lambda}^{(1)}(\kappa r)Y_{\lambda \mu}(\theta, \phi) \right] d\tau
\end{split}
\eeq
In the above formulas $\kappa = \Delta E/c$ with $\Delta E=8.28$ eV being the energy of the isomeric state, and $h_{\lambda}^{(1)}(\kappa r)$ is the spherical Hankel function of the first kind. 
By introducing the reduced nuclear transition probabilities 
\beq
\begin{split}
B(\mathcal{T}\lambda; I_i\rightarrow I_f) &= \frac{1}{2I_i+1}\\
\times \sum_{M_f M_i \mu}&|\la I_fM_f|\mathcal{M} (\mathcal{T}\lambda,\mu)| I_iM_i\ra|^2,
\end{split}
\eeq
the differential cross section becomes
\beq\label{eq_dif_cross2}
\begin{split}
\frac{d\sigma}{d\Omega}= & \frac{4E_fE_i}{c^4}\frac{p_f}{p_i}\\
&\times\sum_{\lambda\mathcal{T}\mu}\left\{\frac{B(\mathcal{T}\lambda; I_i\rightarrow I_f)}{(2\lambda+1)^3}
\frac{1}{2} \sum_{\nu_i \nu_f} |\la \phi_f |\mathcal{N} (\mathcal{T}\lambda,\mu)|\phi_i\ra|^2\right\}
\end{split}
\eeq

Note that we have not specified the detailed forms of the electronic states $|\phi_i\ra$ and $|\phi_f\ra$. As will be shown below, the choice of the electron wave function has big effects on the nuclear excitation cross section. In the following part of this section, we will show forms of excitation cross sections with different choices of electron wave functions, such as Schr\"odinger distorted waves (DWs), Dirac DWs, and Dirac plane waves (PWs).

\subsection{Schr\"odinger DWs} 
Schr\"odinger DWs $|\phi\ra \equiv |\bm{k}\ra^{(\pm)}$ are eigenstates of the time-independent Schr\"odinger equation
\beq
\left[-\frac{\hbar^2}{2m}\nabla^2+V(r) \right]|\bm{k}\ra^{(\pm)} =E|\bm{k}\ra^{(\pm)} ,
\eeq
and they can be expanded into partial-wave series \cite{alder1958errata, landau2013quantum}
\beq\label{eq_SDPWs}
|\bm{k}\ra^{(\pm)}=\sum_{lm} \frac{4\pi}{k}(-1)^mi^le^{\pm i d_{El}}Y_{l,-m}(\hat{\bm{k}})Y_{l,m}(\hat{\bm{r}}) R_l(kr).
\eeq
The initial state (before scattering) takes the plus sign and the final state (after scattering) takes the minus sign: $|\phi_i\ra = |\bm{k}_i\ra^{(+)}$ and $|\phi_f\ra = |\bm{k}_f \ra^{(-)}$ \cite{landau2013quantum}. The total phase shift $d_{El} = \delta_{El} + \Delta_{l}$ with $\delta_{El}$ being the inner phase shift and $\Delta_{l}$ being the Coulomb phase shift, and $R_l(kr)$ is the radial wave function.

The Schr\"odinger equation does not contain electron spin, so Eq. \eqref{eq_dif_cross2} becomes
\beq\label{eq_dif_cross3}
\begin{split}
\frac{d\sigma}{d\Omega}= & \frac{4E_fE_i}{c^4}\frac{p_f}{p_i}\\
&\times\sum_{\lambda\mathcal{T}\mu}\left\{\frac{B(\mathcal{T}\lambda; I_i\rightarrow I_f)}{(2\lambda+1)^3}
|\la \phi_f |\mathcal{N} (\mathcal{T}\lambda,\mu)|\phi_i\ra|^2\right\}.
\end{split}
\eeq
In the non-relativistic limit the multipole transition operators of the scattering electron in Eqs. (\ref{eq_el_matrix_E}-\ref{eq_el_matrix_M}) take simplified forms
\beqa
\mathcal{N} (E \lambda,\mu) &=& - \frac{1}{r^{\lambda+1}} Y_{\lambda \mu}(\theta,\phi) ~,\\
\mathcal{N} (M \lambda,\mu) &=& -\frac{1}{c\lambda}\bm{L}\cdot \nabla
\left[ \frac{1}{r^{\lambda+1}} Y_{\lambda \mu}(\theta,\phi) \right] .
\eeqa
With these non-relativistic simplifications, the total excitation cross section can be expressed as
\beq\label{eq_SCross_E}
\begin{split}
\sigma_{E\lambda}=&\frac{64{\pi}^2}{(2\lambda+1)^2}\frac{k_i k_f}{v_i^2} B(E\lambda,I_i\rightarrow I_f) ~~~~~~~~~~~~~~~~\\
&\times\sum_{l_il_f}(2l_i+1)(2l_f+1)\\
&\times\left(\medspace\begin {array} {ccc} {l_i} & {l_f} & {\lambda} \\ {0} & {0} & {0} \\\end {array}\medspace\right)^2|M_{l_il_f}^{\lambda+1}|^2 ~,\\
\end{split}
\eeq
and
\beq\label{eq_SCross_M}
\begin{split}
\sigma_{M\lambda}=&\frac{64{\pi}^2(\lambda+1)}{\lambda(2\lambda+1)}\frac{v_f}{v_ic^2}B(M\lambda,I_i\rightarrow I_f)\\
&\times\sum_{l_il_f}(2l_i)^2(l_i+1)(2l_i+1)(2l_f+1)\\
&\times\left(\medspace\begin {array} {ccc} {l_i+1} & {l_f} & {\lambda} \\ {0} & {0} & {0} \\\end {array}\medspace\right)^2
\left\{\medspace\begin {array} {ccc} {\lambda} & {\lambda} & 1 \\ {l_i} & {l_i + 1} & {l_f} \\\end {array}\medspace \right\}^2\\
&\times|M_{l_il_f}^{\lambda+2}|^2 ~.
\end{split}
\eeq
In the above equations the radial matrix element $M$ is defined by
\beq\label{eq_SRadial}
M_{l_il_f}^{\lambda+1}=\frac{1}{ k_i k_f}\int_0^\infty \left[R_{l_f}(k_fr) \frac{1}{r^{\lambda+1}} R_{l_i}(k_ir)\right]r^2 \ dr ~,
\eeq
and $M_{l_il_f}^{\lambda+2}$ is similarly defined by replacing $r^{\lambda+1}$ by $r^{\lambda+2}$.

\subsection{Dirac DWs}

Dirac DWs $|\phi\ra \equiv |\bm{k}\nu \ra^{(\pm)}$ are eigenstates of the time-independent Dirac equation
\beq
\left[ -ic\bm{\alpha}\cdot \nabla+\beta c^2 +V(r) \right]|\bm{k}\nu \ra^{(\pm)}=E|\bm{k}\nu \ra^{(\pm)} ~,
\eeq
and they can be expanded into partial-wave series \cite{rose1961relativistic, berestetskii1982quantum}
\beq\label{eq_DDPWs}
\begin{split}
|\bm{k}\nu \ra^{(\pm)}&=\frac{4\pi}{k}\sqrt{\frac{E+m_ec^2}{2E}}\\
&\times\sum_{\eta m}\left[\Omega_{\eta m}^*(\hat{\bm{k}})\chi^{\nu} \right]e^{\pm i d_{E\eta}}
\binom{g_{\eta}(r)\Omega_{\eta m}(\hat{\bm{r}})}{-if_{\eta}(r)\Omega_{-\eta m}(\hat{\bm{r}})}~.
\end{split} 
\eeq
The initial state (before scattering) takes the plus sign and the final state (after scattering) takes the minus sign: $|\phi_i \ra = |\bm{k}_i \nu_i \ra^{(+)}$ and $|\phi_f \ra = |\bm{k}_f \nu_f \ra^{(-)}$. The quantum number $\eta$ is given by
\beq
\eta=(l-j)(2j+1).
\eeq
For $\eta<0$, $l$ is changed to $l'=2j-l$. 
$\Omega_{\eta m}$ are spherical spinors
\beq
\Omega_{\eta m}\equiv \Omega_{jlm}=\sum_{\nu=\pm 1/2 }\la l,1/2,j | m-\nu,\nu,m \ra \chi^{\nu}~.
\eeq
The total phase shift $d_{E\eta}=\delta_{E\eta}+\Delta_{\eta}$, with $\delta_{E\eta}$ and $\Delta_{\eta}$ being the inner phase shifts and the Dirac Coulomb phase shifts, respectively. 
$g_{\eta}(r)$ and $f_{\eta}(r)$ are radial wave functions.

The total excitation cross section is given by
\beq\label{eq_DCross}
\begin{split}
\sigma_{\mathcal{T}\lambda}&=\frac{8\pi^2}{c^4}\frac{p_f}{p_i}\frac{E_f+m_ec^2}{p_f^2}\frac{E_i+m_ec^2}{p_i^2}\\
&\times\sum_{l_i,j_i,l_f,j_f}\frac{\kappa^{2\lambda+2}}{(2\lambda-1)!!^2}B(\mathcal{T} \lambda,I_i\rightarrow I_f)\\
&\times\frac{(2l_i+1)(2l_f+1)(2j_i+1)(2j_f+1)}{(2\lambda+1)^2}\\
&\times\left(\medspace\begin {array} {ccc} {l_f} & {l_i} & {\lambda} \\ {0} & {0} & {0} \\\end {array}\medspace\right)^2\left\{\medspace\begin {array} {ccc} {l_i} & {\lambda} & {l_f} \\ {j_f} & {1/2} & {j_i} \\\end {array}\medspace\right\}^2\\
&\times|M_{fi}^{\mathcal{T}\lambda }|^2,
\end{split}
\eeq
where the radial matrix element $M_{fi}^{\mathcal{T}\lambda }$ is given by
\beq\label{eq_DRadial}
\begin{split}
M_{fi}^{E\lambda}=&\int_0^{\infty}\left\{h_{\lambda}^{(1)}(\kappa r)\left[g_i(r)g_f(r)+f_i(r)f_f(r)\right]r^2 \right.\\
&\left. -\frac{\kappa}{\lambda}h_{\lambda-1}^{(1)}(\kappa r)\left[g_i(r)g_f(r)+f_i(r)f_f(r)\right]r^3 \right\}dr,\\
M_{fi}^{M\lambda}=&\frac{\eta_i+\eta_f}{\lambda} \\
\times &\int_0^{\infty}h_{\lambda}^{(1)}(\kappa r)\left[g_i(r)f_f(r)+g_f(r)f_i(r)\right]r^2dr.
\end{split}
\eeq
For $\kappa r\ll 1$ the asymptotic form of the spherical Hankel function $h_{\lambda}^{(1)}(\kappa r)\approx -i(2\lambda-1)!!/(\kappa r)^{\lambda+1}$ \cite{abramowitz1988handbook} may be used.

\subsection{Dirac PWs}

When the potential from the ion core is neglected, the Dirac DWs reduce to Dirac PWs with the simple form 
$|\phi\ra =|u\ra^{\nu} e^{i \bm{k}\cdot \bm{r}}$, where $|u\ra^{\nu}$ is the four-component spinor with positive energy
\beq
| u \ra^{\nu} = 
\sqrt{\frac{E+m_ec^2}{2E} } 
\begin{pmatrix}
    \chi^{\nu} \\ \frac{\bm{\sigma} \cdot \bm{p} c  }{E+m_ec^2} \chi^{\nu}
\end{pmatrix} .
\eeq
The form of $\chi^{\nu}$ is given in Eq. (\ref{e.chinu}).
It can be shown that the differential excitation cross sections can be written in the following concise forms
\beq\label{eq_PWCross}
\begin{split}
\frac{d\sigma_{E2}}{d\Omega}&=\frac{2\pi}{75c^2}B(E2,I_i\rightarrow I_f)\frac{K^4}{k_i^2}(V_T+\frac{2}{3}V_L)~,\\
\frac{d\sigma_{M1}}{d\Omega}&=\frac{8\pi}{9c^2}B(M1,I_i\rightarrow I_f)\frac{K^2}{k_i^2}V_T~,
\end{split}
\eeq
where $\bm{k}_{i/f}$ is the initial (final) wave vector of the electron, and $\bm{K}=\bm{k}_i-\bm{k}_f$ is the momentum transfer in the scattering process. $V_T$ and $V_L$ are wave vector-dependent functions 
\beq\label{eq_PWVTVL}
\begin{split}
&V_T=k_ik_f\frac{(k_i^2+k_f^2-\kappa^2)^2K^2-2(\bm{k}_i\cdot \bm{K})(\bm{k}_f\cdot \bm{K})}{K^2(K^2-\kappa^2)^2},\\
&V_L=k_ik_f\frac{2k_i^2+2k_f^2+4c^2-\kappa^2-K^2}{K^4}.
\end{split}
\eeq
These formulas have also been given in \cite{alder1958errata}. The total cross section is calculated by integrating the differential cross section over the solid angle.

\subsection{Calculation of radial wave functions}

The calculation of the excitation cross sections eventually reduces to the calculation of the radial wave functions and the radial matrix elements [Eqs. \tcb{\eqref{eq_SRadial}} and \tcb{\eqref{eq_DRadial}}]. In this article the radial wave functions are calculated using the code RADIAL \cite{salvat2019radial}. The electron density distribution $\rho(r)$ of the $^{229}$Th atom or ion is calculated using a Dirac-Hartree-Fork-Salter method \cite{liberman1965self,liberman1971relativistic}, and the potential from the electron cloud can be calculated as
\beq
V_\text{el} (r) = \int_0^\infty \frac{\rho(r^{'})}{|\bm{r}-\bm{r}^{'}|}d\bm{r}^{'} ~.
\eeq

The charge density of the nucleus is modeled by a Fermi charge distribution
\beq
\rho_n(r)=\frac{\rho_0}{\exp[(r-R_n)/z]+1} ~,
\eeq
where $R_n=1.07 A^{1/3}$ fm with $A$ the mass number, and $z$=0.546 fm. $\rho_0$ is a constant, which equals twice the proton density at $r=R_n$, and is to be determined by a normalization condition \cite{hahn1956high}. 
The potential energy from the nucleus is given by
\beq
V_\text{nuc} (r) = -\int_0^\infty \frac{\rho_n(r^{'})}{|\bm{r}-\bm{r}^{'}|}d\bm{r}^{'} ~.
\eeq
The total potential felt by the scattering electron is $V(r) = V_\text{nuc} (r) + V_\text{el} (r)$.

\section{Numerical results and discussions}\label{III}

In this section we present numerical results and analyze relevant elements affecting the nuclear excitation cross section. The importance of ion-core potentials on the excitation cross section can be seen by comparing cross sections calculated using PWs versus DWs. The importance of relativistic effects on the excitation cross section can be seen by comparing cross sections calculated using Dirac DWs versus Schr\"odinger DWs. The dependency of excitation cross sections on different ionic states is also checked. Effect of uncertainties in the knowledge of the reduced nuclear transition probabilities on the excitation cross section will be discussed. Although we mainly focus on cross sections with low electron energies, we also show results with a much larger energy range to see how the difference between the DW-cross section and the PW-cross section shrinks and disappears with increasing electron energy. Some further remarks on possible schemes utilizing electronic excitation of the $^{229}$Th nucleus will be given at the end of this section.

\subsection{DWs vs. PWs}\label{III.A}

\begin{figure}
    \centering
    \includegraphics[width=8cm]{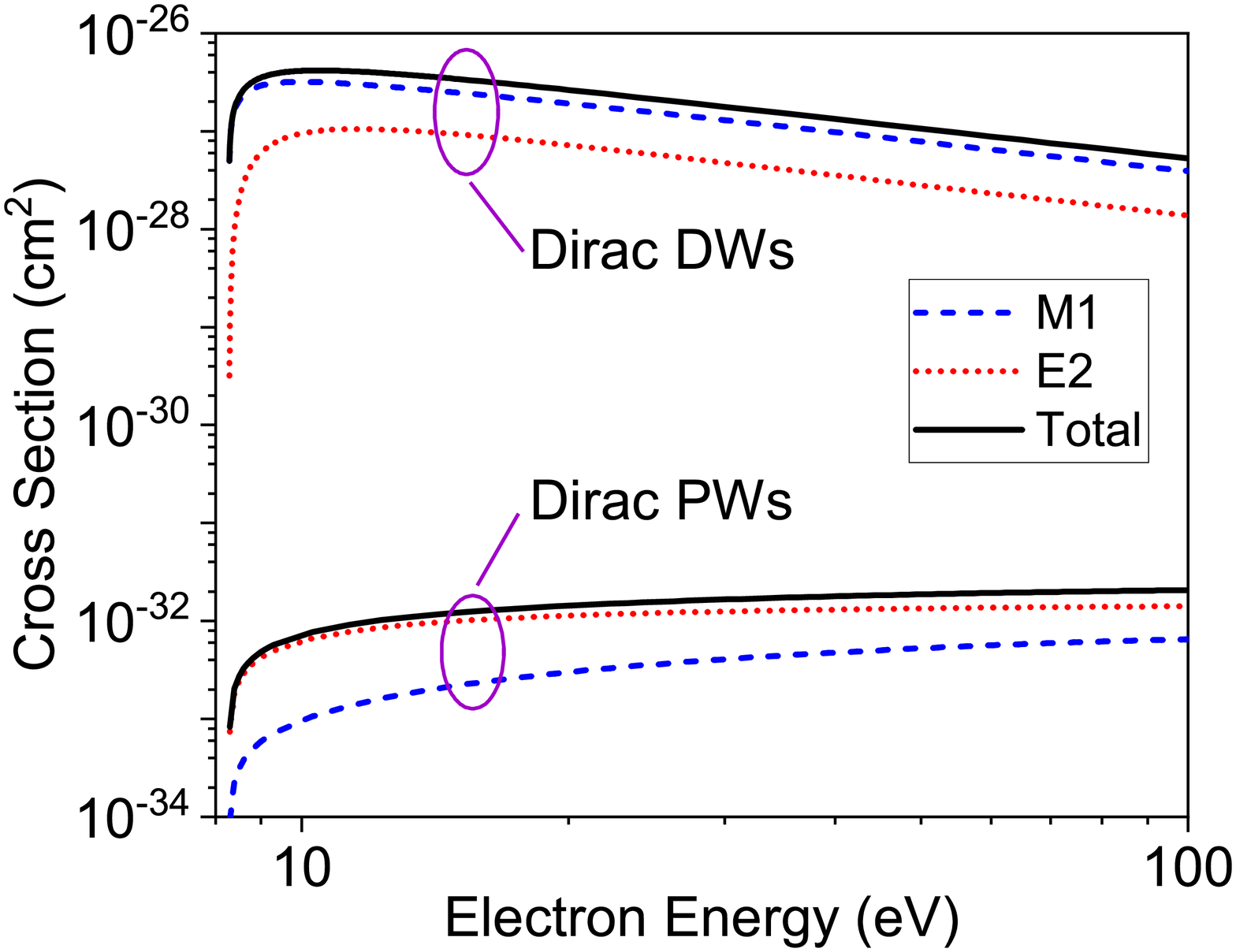}
    \caption{(color line) Isomeric excitation cross section of $^{229}$Th calculated using Dirac DWs and Dirac PWs, as labeled on figure. For each case, the total cross section as well as separated contributions from the $M1$ or the $E2$ channels are shown.}\label{fig.pw}
\end{figure}

Figure \tcb{\ref{fig.pw}} shows the nuclear isomeric excitation cross sections with Dirac DWs and Dirac PWs. For each case, separated contributions from the $E2$ and the $M1$ channels are shown, in addition to the total cross section. One sees that the DWs lead to cross sections about 5 to 6 orders of magnitude higher than the PWs do. For the DW case, the cross sections are on the order of $10^{-27}$ to $10^{-26}$ cm$^2$ for the energy range shown, whereas for the PW case, the cross sections are on the order of $10^{-33}$ to $10^{-32}$ cm$^2$. Therefore wave function distortion has an extremely large effect on the nuclear excitation cross section.

\begin{figure}
    \centering
    \includegraphics[width=8cm]{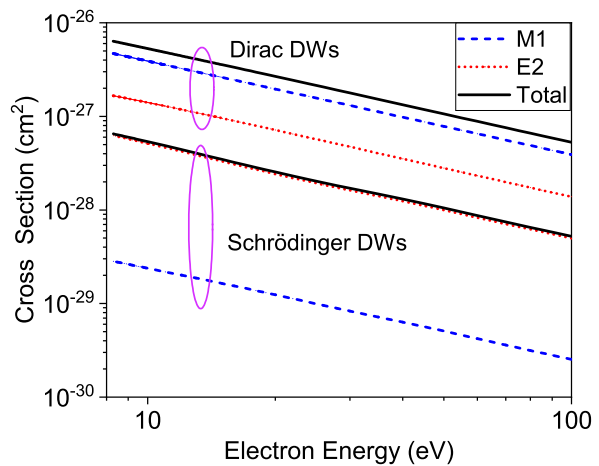}
    \includegraphics[width=8cm]{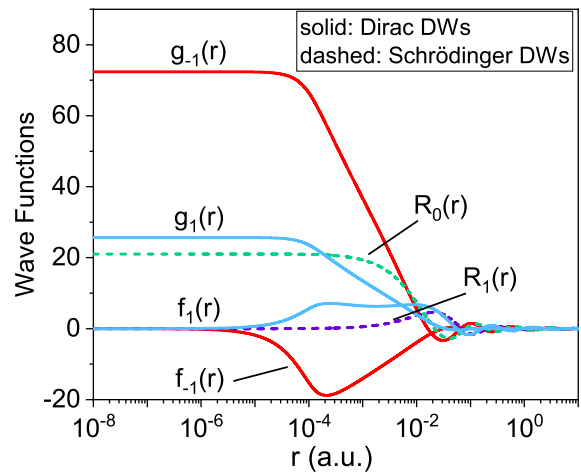}
    \caption{(color line) (Upper) Isomeric excitation cross section of ${^{229}}$Th$^{1+}$ with Dirac DWs and Schr\"odinger DWs, as labeled. Both the total cross section and separated contributions from the $M1$ and the $E2$ channels are shown. (Lower) Radial wave functions of the scattering electron with energy $E$ = 8.3 eV in the potential of the Th$^{1+}$ ion. Only the first few partial waves with relatively large amplitudes are shown. Solid curves are Dirac DWs with $\eta = -1,1$, and dashed curves are Schr\"odinger DWs with $l = 0,1$.}\label{3.relaeff}
\end{figure}

In addition to the overall magnitude, other differences can also be seen: (i) The DW cross sections show a quick increase from the threshold energy of 8.28 eV to about 10 eV, then a slow decrease for energies above 10 eV. In contrast, the PW cross sections show an overall increase. (ii) For the DW case, the $M1$ channel is higher than the $E2$ channel, whereas for the PW case, the $E2$ channel is higher than the $M1$ channel.

It should be noted that both the DW and the PW results shown in Fig. \tcb{\ref{fig.pw}} are calculated using Eq. \tcb{\eqref{eq_DCross}}. PW results are calculated by simply putting $V(r)=0$ in the calculation. We have checked that the PW results are identical to those obtained directly from the analytical formulas of Eqs. \tcb{\eqref{eq_PWCross}} and \tcb{\eqref{eq_PWVTVL}}, as would be expected. Note that the cross section formulas given by Tkalya \cite{tkalya2020excitation} are larger by an overall factor of two. They do not agree to the analytical formulas in the limit $V(r) \rightarrow 0$.

\subsection{Dirac DWs vs. Schr\"odinger DWs}

Cross sections (total, as well as separated contributions from the $E2$ or the $M1$ channels) calculated using Dirac DWs and Schr\"odinger DWs are compared in Fig. \tcb{\ref{3.relaeff}}.  One can see that Dirac DWs lead to excitation cross sections about an order of magnitude higher than Schr\"odinger DWs do. This difference is due to relativistic effects, which turn out to be quite remarkable.

\begin{figure}
    \centering
    \includegraphics[width=8cm]{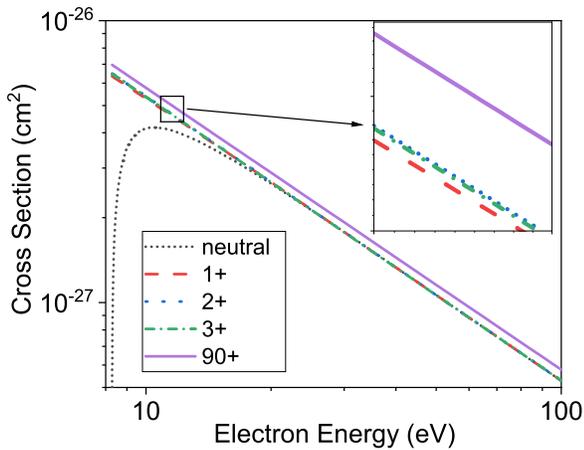}
    \caption{(color line) Isomeric excitation cross section of $^{229}$Th for different ionic states, as labeled. The inset zooms in a small energy range between 11 and 12 eV.}\label{fig.vle}
\end{figure}

This result might seem to be rather unexpected at first glance, since the electron energies considered here are low (below 100 eV). The remarkable relativistic effect results from two reasons. One is that $^{229}$Th has a large $Z = 90$. The other is that it is actually the wave function very close ($r < 10^{-2}$ a.u.) to the nucleus that contributes dominantly to the radial integrals in Eqs. \tcb{\eqref{eq_SRadial}} and \tcb{\eqref{eq_DRadial}}. At these distances the Dirac DWs and the Schr\"odinger DWs have very different magnitudes, as shown in the lower panel of Fig. \tcb{\ref{3.relaeff}}. Compared to the non-relativistic wave functions, the relativistic wave functions have much lager amplitudes for $r<10^{-2}$ a.u. 

One also notices from the upper panel of Fig. \tcb{\ref{3.relaeff}} that for the Dirac case, the $M1$ channel is several times higher than the $E2$ channel, while for the Schr\"odinger case, the $E2$ channel is over an order of magnitude higher than the $M1$ channel.

\subsection{Effect of Ionic States}

Figure \tcb{\ref{fig.vle}} displays the nuclear excitation cross sections for different ionic states, namely, the neutral Th atom, the Th$^{1+}$, Th$^{2+}$, Th$^{3+}$, and Th$^{90+}$ ions. 

The overall feature is that the nuclear excitation cross section depends rather weakly on the ionic state. The cross sections for the ionic cases are very close to each other throughout the whole energy range. A minor exception is the neutral case for energies below about 10 eV. The reason for this weak ionic-state dependency is that only the electron wave function very close ($r < 10^{-2}$ a.u.) to the nucleus contributes dominantly to nuclear excitation, and almost all electrons are outside this radius. In rare circumstances, however, the electron cloud may affect the wave function and the cross section more appreciably, as in the case of the neutral atom. Similar results have also been reported in Ref. \cite{tkalya2020excitation}.

\subsection{Uncertainties in the reduced nuclear transition probabilities}

\begin{figure}
    \centering
    \includegraphics[width=8cm]{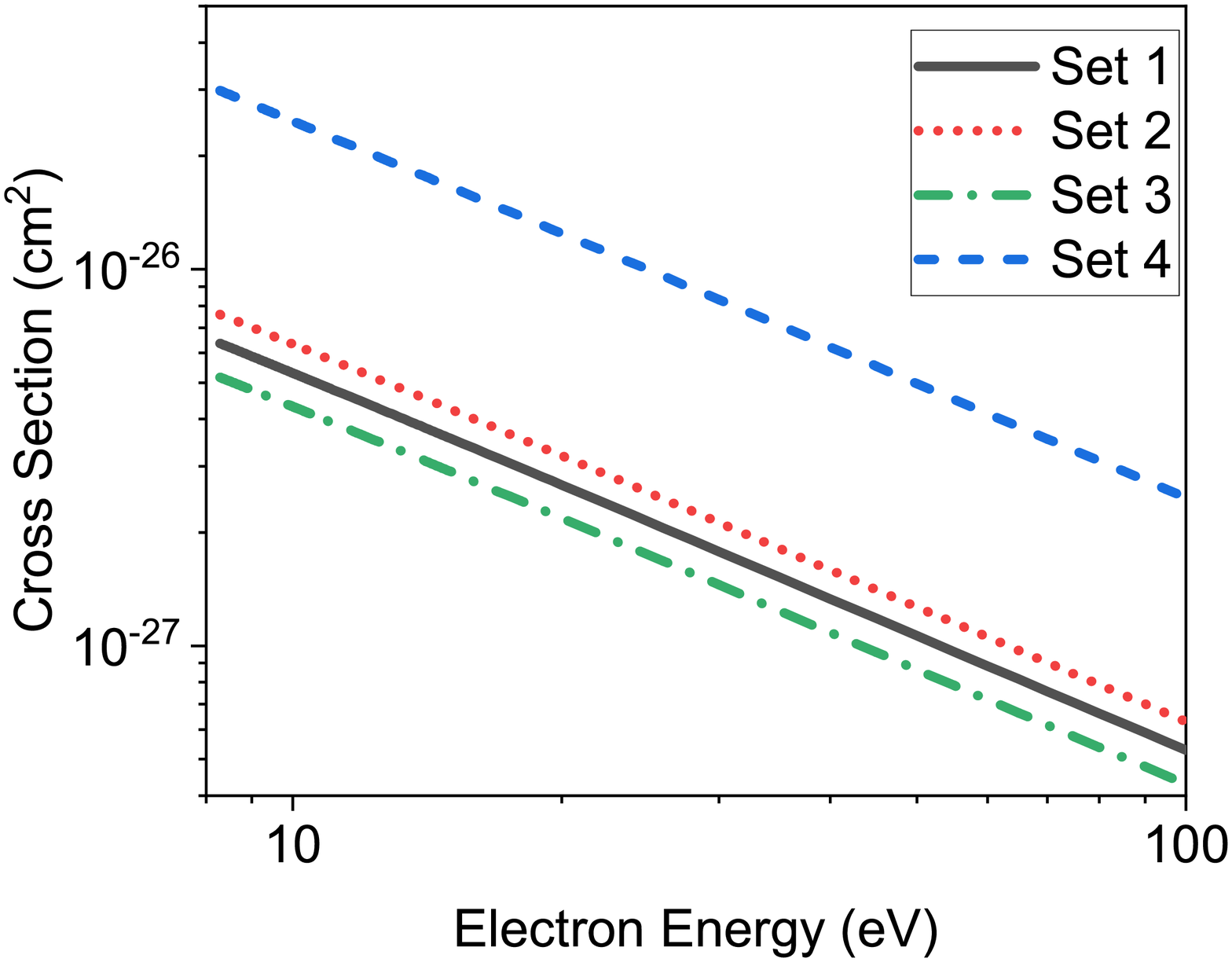}
    \caption{(color line) Isomeric excitation cross section of $^{229}$Th$^{1+}$ for four different sets of the reduced nuclear transition probabilities, as labeled. See text for the detailed values of the reduced transition probabilities.}\label{fig.3sets}
\end{figure}

The reduced nuclear transition probabilities $B(E2/M1)$ are determined either from nuclear model calculations \cite{gulda2002nuclear}, for example in the framework of a quasiparticle-phonon model with inclusion of Coriolis couplings \cite{blin1977theory, aas1996enhanced}, or from experimental data analyses \cite{bemis1988coulomb, gulda2002nuclear, barci2003nuclear, ruchowska2006nuclear} exploiting Alaga rules \cite{tkalya2015radiative, dykhne1998229m}. In the calculations above we have used the values suggested by Minkov and P\'alffy in 2017 \cite{minkov2017reduced}, and this set of values is denoted as set 1, as given below. Here we use other suggested values for the purpose of comparison. For example, in 2021 Minkov and P\'alffy suggested an updated range of values \cite{minkov2021th}, and we denote the upper and lower limits of the range as set 2 and set 3. Set 4 is from experimental data analyses \cite{dykhne1998229m, tkalya2020excitation}. These sets of reduced transition probabilities are listed as follows (W.u. means Weisskopf units):
\beqa
\text{Set } 1: B(E2, e\rightarrow g) &=& 27\ \rm{W.u.} \nonumber\\ 
              B(M1, e\rightarrow g) &=& 0.0076\ \rm{W.u.} \nonumber\\
\text{Set } 2: B(E2, e\rightarrow g) &=& 42.9\ \rm{W.u.} \nonumber\\
              B(M1, e\rightarrow g) &=& 0.008\ \rm{W.u.} \nonumber\\
\text{Set } 3: B(E2, e\rightarrow g) &=& 33.8\ \rm{W.u.} \nonumber\\
              B(M1, e\rightarrow g) &=& 0.005\ \rm{W.u.} \nonumber\\
\text{Set } 4: B(E2, e\rightarrow g) &=& 17.55\ \rm{W.u.} \nonumber\\
              B(M1, e\rightarrow g) &= &0.048\ \rm{W.u.} \nonumber
\eeqa

{\it Ab initio} calculations of $B(E2/M1)$ for a nucleus like $^{229}$Th are out of reach in the foreseeable future, and there is no conclusive means to judge which set of values is better than other sets. This is the current status of knowledge about the reduced nuclear transition probabilities.
Uncertainties in the values of the reduced transition probabilities lead to uncertainties in the nuclear excitation cross sections within an order of magnitude, as shown in Fig. \tcb{\ref{fig.3sets}}. Only the total cross section for the $^{229}$Th$^{1+}$ ion is shown for each set, and the electron wave functions are Dirac DWs. 

Note that the direction of nuclear transition also matters
\beqa
 \frac{B(E2/M1; g \rightarrow e)}{B(E2/M1; e \rightarrow g)} = \frac{2 I_e + 1}{2 I_g + 1} = \frac{2}{3}~.
\eeqa
In Ref. \cite{tkalya2020excitation} there seems to be a confusion in the transition direction of set 1.

\subsection{Cross section with higher electron energies}

\begin{figure}
    \centering
    \includegraphics[width=8cm]{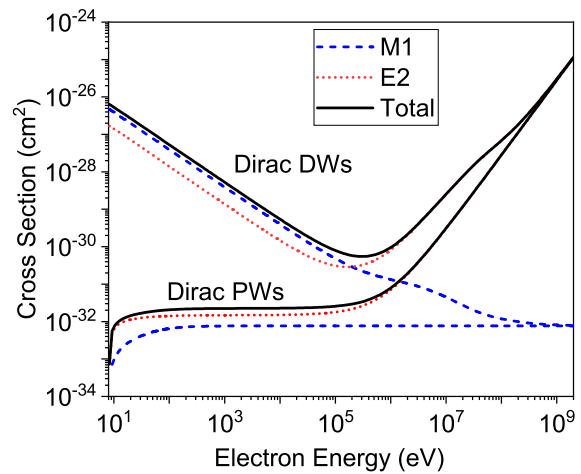}
    \caption{(color line) Isomeric excitation cross section of $^{229}$Th$^{1+}$ with Dirac DWs and PWs, as labeled, for an extended energy range. Separated contributions from the $E2$ or the $M1$ channels are also shown. }\label{fig.highE}
\end{figure}

We have mainly focused on excitation cross sections with electron energies below 100 eV. In Fig. \tcb{\ref{fig.highE}}  we present excitation cross sections with a much larger energy range up to $10^9$ eV, or 1 GeV. We show cross sections from both the Dirac DWs and the Dirac PWs.

Let us look at the total cross sections. One can see that the DW cross section drops first with the electron energy, reaches a minimum around 0.5 MeV, and then increases with the electron energy. In contrast, the PW cross section increases quickly with the electron energy below 10 eV, reaches a plateau, then increases quickly with energy above 0.5 MeV. The gap between the DW cross section and the PW cross section shrinks with the increase of the electron energy, as would be expected, although the two agree with each other only with very high electron energies approaching 1 GeV. 

For electron energies below about 0.5 MeV, both the $M1$ and the $E2$ channels contribute, although the $M1$ channel is more important for the DW case and the $E2$ channel is more important for the PW case. For electron energies higher than 0.5 MeV, however, almost all the contributions come from the $E2$ channel, for both cases. 

Note that the purpose of Fig. \tcb{\ref{fig.highE}} is just to see the comparison between the DW and the PW cross sections. Only the nuclear ground state and the isomeric excited state are considered. Excitation to higher nuclear excited states is not taken into account here.

\subsection{Further remarks}

We consider here a few schemes in which electronic excitation of the $^{229}$Th nucleus may be implemented. The most straightforward scheme is to use an external electron beam with electron energies tuned to values corresponding to the highest excitation cross sections, e.g. below 100 eV.

An alternative scheme is to use the $^{229}$Th atom's own electrons. The idea is to use a strong laser pulse to pull out one or several electrons from the $^{229}$Th atom (i.e. strong-field ionization), and then to drive the electron(s) back to collide with and excite the $^{229}$Th nucleus. This process is called recollision \cite{Kulander-93, Schafer-93, Corkum-93} which is the core process of strong-field atomic physics. The energy of the recolliding electron is usually several tens of eV, which is precisely the energy region with the highest nuclear excitation cross sections. We proposed this scheme in Refs. \cite{wang2021exciting, wang2021strong} with extended calculation results presented in Ref. \cite{Wang2022}.

Another scheme is to use a plasma environment. In thermal equilibrium the electron has a Maxwell-Boltzmann distribution that will be integrated to calculate the nuclear excitation rate. A particularly promising approach is to start from an atomic cluster, and to use a strong laser pulse to interact with the cluster \cite{feng2022femtosecond}. The atoms in the cluster are ionized releasing electrons, which can be confined in the cluster for a time scale on the order of 1 ps and excite the nucleus. The cluster has a solid-state atomic density, so the flux density of the electron can be much higher than that in a typical plasma.

\section{Conclusion}\label{IV}

In this article we consider nuclear isomeric excitation of $^{229}$Th via inelastic electron scattering. A theoretical framework is presented for the calculation of the excitation cross section on the level of Dirac distorted wave Born approximation. Numerical results are shown with detailed analyses on elements that affect the excitation cross section. We show how the excitation cross section changes if the ion-core potential is removed (DWs vs. PWs), if the relativistic effect is removed, if the ionic state is changed, and if the reduced nuclear transition probabilities are changed.
Special emphases are given to low electron energies with which the cross sections are relatively high. Nevertheless, we also show cross sections with an extended energy range for the curiosity of comparing PW cross sections with DW cross sections.

$\newline$

We acknowledge funding support from Science Challenge Project of China No. TZ2018005, NSFC No. 12088101, and NSAF No. U1930403.

\end{document}